\documentclass[a4paper]{jpconf}
\usepackage{graphicx}
\begin{document}
\title{Impact of error analysis on the composition the outer crust of a neutron star}

\author{D Neill, K Medler, A Pastore and C Barton }
\address{Department of Physics, University of York, Heslington, York, Y010 5DD, UK}
\ead{alessandro.pastore@york.ac.uk}
\ead{charles.barton@york.ac.uk}

\begin{abstract}
By means of bootstrap technique, we perform a full error analysis on the Duflo-Zucker mass model. We illustrate the impact of such  study on the predicted chemical composition of the outer crust of a non-accreting neutron star.
We define an existence probability for each nuclear species as a function of the depth of the crust. We observe that, due to statistical uncertainties, instead of having a well defined transition between two successive layers, we have a mixture of two species.
\end{abstract}

\section{Introduction}
The study of neutron stars (NS)  is important for our understanding of nuclear matter in the most extreme conditions.  The recent discovery  of gravitational waves~\cite{abbott2017gw170817} has opened an entire new way of studying such objects, providing us with new set of observables.

The matter within the NS arranges in layers with different properties due to the large pressure gradient~\cite{chamel2008physics}. Going from the outside toward the inside (low density to high density), we find the \emph{envelope}: a thin layer of iron atoms. At densities of $\rho\approx 10^4$ $g/cm^3$, the outer crust begins. Here a gas of (ultra) relativistic  electrons surrounds ionised nuclei forming a crystal lattice.
Throughout the \emph{outer crust} the fermi energy of the electrons increases with density, thus making it energetically favourable for electron capture to occur. The neutronisation of matter leads to the production of very neutron rich nuclei. The process continues until the drip-line is reached and neutrons start dripping off  forming a uniform gas~\cite{pastore2011superfluid,pastore2012superfluid,chamel2012neutron,pearson2012inner,pastoreDRIP,pastore2017new,pas17b}. This region of the star is called the \emph{inner crust}.
In the following layer, the \emph{core}, the density is high enough so that nuclei dissolve into a Fermi liquid. 
The composition of the core is still matter of open debate and several models concerning its structure have been suggested over the years~\cite{alford2007astrophysics,sharma2015unified,chatterjee2016hyperons,vidana2018d}.

 In this article, we focus our attention on the outer crust region and how a complete error analysis impacts the prediction of the equation of state.
To determine the sequence of nuclei within the crust, we follow the method suggested by Baym, Pethick and Sutherland (BPS) in 1971~\cite{baym1971ground}: for each input pressure value and given lattice configuration, one gets the number of neutrons (N) and protons (Z) that minimises the Gibbs free energy.
A key ingredient in these calculations is the determination of the nuclear binding energy of nuclei close to the neutron drip line. For simplicity, we will ignore in the following the presence of binary or ternary ionic compounds~\cite{cha16}.

Since most of the nuclei predicted are extremely neutron rich, they are not yet accessible experimentally and thus one needs to use a theoretical model. To predict nuclear masses, we can find several models within the literature and they all rely on some adjustable parameters~\cite{sob14}.
In the present article, we want to investigate the role of the statistical errors associated with these parameters on the determination of the chemical composition of the NS.

The article is organised as follows: in Sec.\ref{sec:DZ} we briefly discuss the main features of the DZ mass model and the statistical procedure used to estimate covariance matrix and error bars. In Sec.\ref{sec:outer}, we  present how the statistical errors on the nuclear binding energies impact the calculations of the outer crust composition. We finally provide our conclusions in Sec.\ref{sec:conc}.

\section{Duflo-Zucker mass model}\label{sec:DZ}
The Duflo-Zucker (DZ) mass model~\cite{duflo1995microscopic}  is a very successful macroscopic model based on few adjustable parameters.
Following the detailed explanations provided in Refs~\cite{mendoza2010anatomy,zuker2011anatomy,qi2015theoretical}, we write the nuclear binding energy (BE) for a given nucleus with N neutrons and Z protons as a sum of ten terms (DZ10) as
\begin{eqnarray}\label{bene}
B_{th}=a_1V_C+a_2 (M+S)-a_3\frac{M}{\rho}-a_4V_T+a_5V_{TS}+a_6s_3-a_7\frac{s_3}{\rho}+a_8s_4+a_{9}d_4+a_{10}V_P\;.
\end{eqnarray}
\noindent We used $A=N+Z$ ; $2T=|N-Z|$  and $\rho=A^{1/3}\left[ 1-\frac{1}{4}\left(\frac{T}{A}\right)^2\right]^2$.
The ten different contributions can be grouped in two categories: in the first one we find terms similar to liquid drop model (LD) as Coulomb ($V_C$), symmetry energy ( $V_T,V_{TS}$) and pairing $V_P$. The other parameters originate from the averaging of shell-model Hamiltonian and they are based on the microscopic single level structure of the nucleus. For a more detailed discussion we refer to Ref.~\cite{mendoza2010anatomy}, where all these terms are described in great detail.

\begin{table}[!h]
\caption{\label{tab:data} Parameters of the DZ10 mass formula given in Eq.\ref{bene}. The quantities are expressed in MeV. The last two columns represent the error on the parameters. See text for details.}
\begin{center}
\lineup
\begin{tabular}{lll}
\br
 &     Value                   & Error   \\
\mr
$a_1$ &  \00.70454     & 0.00037\\
$a_2$ &   17.740     &0.0068   \\
$a_3$ &   16.242& 0.023 \\
$a_4$ &   37.497   & 0.042  \\
$a_5$ &   53.56     & 0.20 \\
$a_6$ &  \00.4573     & 0.0069  \\
$a_7$ &   \02.072     & 0.035 \\
$a_8$ &   \00.0210 & 0.0002  \\
 $a_9$ &  41.41      & 0.21  \\
 $a_{10}$ &  \06.162 &0.088   \\
\br
\end{tabular}
\end{center}
\end{table}

%
\noindent As a first step of our analysis, we have decided to perform  a full fit the DZ mass model using the nuclear mass table of 2012~\cite{wang2012ame2012} and adopting the Non  Parametric Bootstrap (NPB) method illustrated in Ref.~\cite{efr86}. It consists in a simple Monte Carlo sampling of the residual of the fitted model to build the empirical distribution of the parameters.
The NPB has been successfully tested as a simple parameter estimator for a liquid drop model~\cite{pastore2019introduction}, here we prove that is also a valuable tool to estimate more complex models in a larger parameter space.
We define a penalty function of the form

\begin{eqnarray}\label{chi2}
\chi^2=\sum_{N,Z} \frac{\left[B_{exp}(N,Z)-B_{th}(N,Z)\right]^2}{\sigma^2(N,Z)}\;.
\end{eqnarray}

In the fit, we  have excluded all nuclei with $A<14$ and we have considered only measured binding energies  with an experimental error lower than $100$ keV. Ignoring the small experimental uncertainties, we set $\sigma^2(N,Z)=1$ MeV$^2$, thus assuming equal weight to all data. In total we considered 2293 nuclei. 

The resulting parameters $\mathbf{a}=\{a_1,a_2,\dots a_{10}\}$ are provided in Tab.\ref{tab:data}. The DZ10 values are compatible (within error bars) with previous analysis illustrated in Ref.~\cite{mendoza2010anatomy} and based on 1995 mass table~\cite{audi19951995} and  the more recent results given in Ref.~\cite{uta17} based on 2012 mass table~\cite{wang2012ame2012}. In the last column of the table, we also report the errors  obtained as 68\% quantile of the parameters distributions obtained using NPB~\cite{efr86}. We refer to Ref.~\cite{pastore2019introduction} for a more detailed explanation of the NPB method. From Tab.\ref{tab:data}, we observe that the statistical errors are quite small. The errors tell us about the stiffness of the $\chi^2$ surface around the minimum. The model is already well constrained and adding extra data would not change the results. 
Of course, having small statistical errors does not imply small systematic errors~\cite{dobaczewski2014error}.

\begin{figure}[!h]
\begin{center}

\includegraphics[width=0.6\textwidth,angle=0]{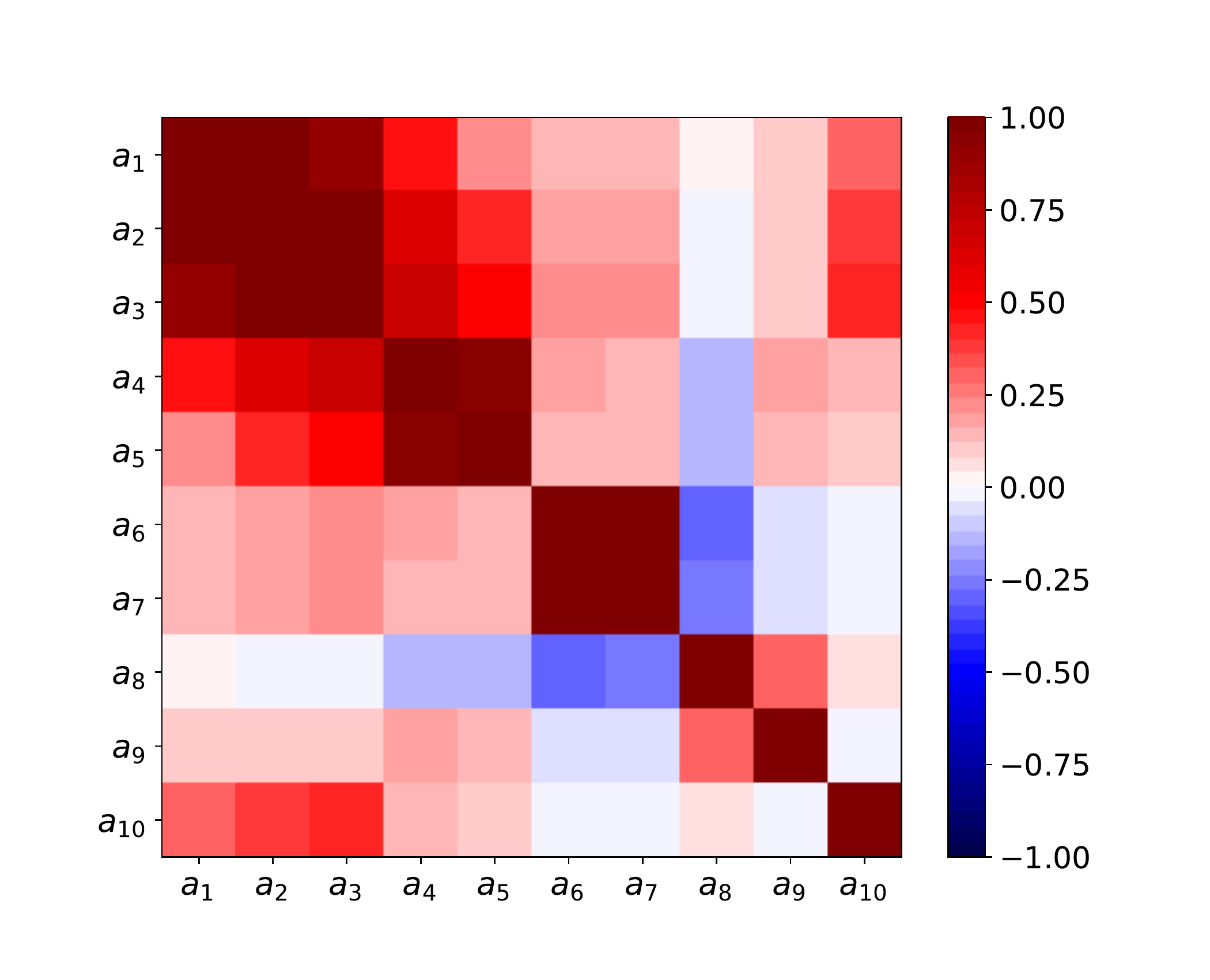}
\end{center}
\caption{(Colour online).Correlation matrix $\mathcal{R}$ obtained using NPB for the parameters of the DZ10 mass model as defined in Eq.\ref{bene}.}
\label{cov}
\end{figure}

As discussed in Ref.~\cite{pastore2019introduction}, it is possible to obtain with NPB  the covariance matrix $\mathcal{C}$ without performing derivatives in parameter space. By rescaling its matrix elements by the variances of individual parameters we obtain  the correlation matrix 
$\mathcal{R}_{ij}=\frac{\mathcal{C}_{ij}}{\sqrt{\mathcal{C}_{ii}}\sqrt{\mathcal{C}_{jj}}}$. The latter contains matrix elements with values in the range $\in[-1,1]$ indicating possible correlations (+1) or anti-correlations (-1) between parameters. The results for the DZ10 model are reported in Fig.\ref{cov}.
We observe that some parameters are strongly correlated among each other. This information may be used to perform a parameter reduction of the model as shown in Ref.~\cite{nik16}.

\noindent The covariance matrix is a key ingredient to propagate errors on the nuclear masses calculated with the DZ10 model~\cite{roca2015covariance}. For each value of masses we thus have an error as $\pm\sqrt{V_{th}(N,Z)}$ where

\begin{eqnarray}\label{err}
V_{th}(N,Z)=\sum_{ij}\frac{\partial B_{th}(N,Z)}{\partial a_i} \mathcal{C}_{ij} \frac{\partial B_{th}(N,Z)}{\partial a_j}\;.
\end{eqnarray} 

\noindent This formula is based on a parabolic approximation of the $\chi^2$ surface around its minimum. We have tested using a full Monte Carlo error propagation the results do not change in a significant way, thus proving that this approximation is working very well in the present case.
We refer to Ref.~\cite{bar89} for more details.

\section{Outer crust}\label{sec:outer}

To determine the proton/neutron fraction within the outer crust, we follow the same procedure presented in Refs~\cite{baym1971ground,ruster2006outer,roca2008impact,pearson2011properties}. 
To take into account the impact of statistical error on the calculations of the outer crust, we have defined a simple Monte Carlo procedure based on a parametric bootstrap (PB)~\cite{kreiss2011bootstrap}.
The method has been already detailed in Ref.~\cite{per14}.
In the case of NPB, one resamples the data without assuming any underlying probability distribution function (PDF); in the case of PB the Monte Carlo samples are extracted by an hypothetical PDF. 
In few words, we generate $10^4$ nuclear mass tables by sampling a Gaussian distribution for each nuclear mass having mean the predicted value of binding energy of the DZ10 model and as variance the theoretical error $V_{th}$ defined in Eq.\ref{err}. 

For each of the mass tables, we apply the minimisation procedure discussed in Ref~\cite{roca2008impact} and determine the chemical composition of the NS crust. 
It is important to mention that whenever an experimental measurement (not an extrapolation) is available, we always replace the theoretical mass  value with the corresponding experimental one, since the typical experimental error is way more accurate.
\noindent As a consequence, in the current analysis, we study the effect of the statistical errors only for values of the pressure greater than $P\approx10^{-4}$ [MeVfm$^{-3}$]. This is roughly the pressure regime where our procedure predicts the existence of nuclei not yet measured experimentally.

\begin{figure}[!h]
\begin{center}

\includegraphics[width=0.6\textwidth,angle=0]{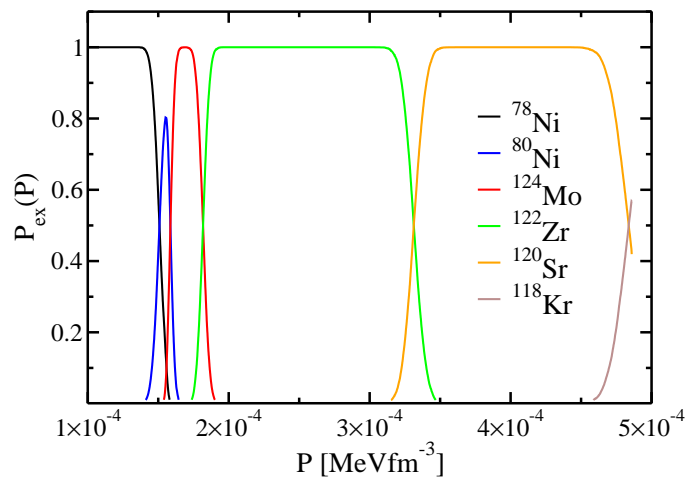}
\end{center}
\caption{(Colour online). Existence probability of various nuclear species as a function of the pressure of the system.}
\label{prob}
\end{figure}

 We finally define an existence probability, $P_{ex}$, for each nuclear species as the ratio of the number of times the species appears at a given pressure in one of the simulated crusts divided by the total number of simulations ($10^4$). $P_{ex}$ can take any value, by construction, in the range $P_{ex}\in[0,1]$.  In Fig.\ref{prob}, we plot the existence probability  of various nuclear species as a function of the pressure of the system.

We observe that there are no sharp transitions between the different layes of the crust, due to the statistical uncertainties of the model. To reduce such errors either more accurate models or experiment are required.
It is interesting to observe that at the edge between the inner and outer crust our prediction give roughly equal probability to i $^{120}$Sr and $^{118}$Zr.
In particular, we notice that the transition outer-inner crust is quite uncertain. Both nuclei $^{120}$Sr and $^{118}$Zr may exist with roughly equal probability.

The direct consequence of our analysis is that the EoS in this region of the crust is no more a sharp line, but we should define it in terms of confidence interval. It is well known that the several mass model predict similar gross features of the EoS of the outer crust, \emph{i.e.} the preference of nuclei around shell closure N=50 and N=82, but the exact nucleus predicted by each model at each depth of the star may be very different~\cite{ruster2006outer}. The systematic use of error bars to 
 
\section{Conclusions}\label{sec:conc}
 
 We have applied the Bootstrap method~\cite{pastore2019introduction} methods to estimate the correlations and errors bars on the parameters of the Duflo-Zucker model~\cite{duflo1995microscopic}. The results obtained with such a method are in good agreement with previous findings based on different statistical methods~\cite{qi2015theoretical}.

By means of Monte Carlo methods, we have suggested a possible methodology to estimate the role of model error bars~\cite{dobaczewski2014error} on the chemical composition of the outer crust of a neutron star.
We have shown that the transition from one nuclear species to the other is not sharp, but due to statistical uncertainties, we identify regions of coexistence.
The extension of such regions crucially depends on the propagated errors and thus on the adopted model.

\ack

This work has been supported by STFC Grant No. ST/P003885/1.

\section*{References}
\bibliographystyle{iopart-num}

\bibliography{biblio}

\end{document}